\documentclass[pre,aps,twocolumn,superscriptaddress,nofootinbib]{revtex4-1}
\usepackage{graphicx}
\usepackage{amssymb,latexsym,amsthm,amsmath}    
\usepackage{mathtools}
\usepackage{float}
\usepackage[draft]{changes}
\usepackage{hyperref}
\usepackage{color}
\usepackage{colortbl}
\usepackage{ulem} % para tachar palabras

\begin{document}
\title{Dynamical complexity as a proxy for the  network degree distribution}

\author{A. Tlaie}
\affiliation{Complex Systems Group \& GISC, Universidad Rey Juan Carlos, 28933 M\'ostoles, Madrid, Spain}
\affiliation{Center for Biomedical Technology, Universidad Polit\'ecnica de Madrid, 28223 Pozuelo de Alarc\'on, Madrid, Spain}
\affiliation{Department of Applied Mathematics and Statistics, ETSIT Aeron\'auticos, Universidad Polit\'ecnica de Madrid, 28040 Madrid, Spain }
\author{I.  Leyva}
\affiliation{Complex Systems Group \& GISC, Universidad Rey Juan Carlos, 28933 M\'ostoles, Madrid, Spain}
\affiliation{Center for Biomedical Technology, Universidad Polit\'ecnica de Madrid, 28223 Pozuelo de Alarc\'on, Madrid, Spain}
\author{R. Sevilla-Escoboza}
\affiliation{Centro Universitario de los Lagos, Universidad de Guadalajara, Jalisco 47460, M\'exico}
\author{V.P. Vera-Avila}
\affiliation{Centro Universitario de los Lagos, Universidad de Guadalajara, Jalisco 47460, M\'exico}
%\author{J.M. Buld\'u}
%\affiliation{Complex Systems Group \& GISC, Universidad Rey Juan Carlos, 28933 M\'ostoles, Madrid, Spain}
%\affiliation{Center for Biomedical Technology, Universidad Polit\'ecnica de Madrid, 28223 Pozuelo de Alarc\'on, Madrid, Spain}
\author{I.  Sendi\~na-Nadal}
\affiliation{Complex Systems Group \& GISC, Universidad Rey Juan Carlos, 28933 M\'ostoles, Madrid, Spain}
\affiliation{Center for Biomedical Technology, Universidad Polit\'ecnica de Madrid, 28223 Pozuelo de Alarc\'on, Madrid, Spain}

\begin{abstract}

  We explore the relation between the topological relevance of a node in a complex network and the individual dynamics it exhibits. When the system is weakly coupled, the effect of the coupling strength against the dynamical complexity of the nodes is found to be a function of their topological  role, with nodes of higher degree displaying lower levels of complexity. We provide several examples of theoretical models of chaotic oscillators, pulse-coupled neurons and experimental networks of nonlinear electronic circuits evidencing such a hierarchical behavior. Importantly, our results imply that it is possible to infer the degree distribution of a network only from individual dynamical measurements.  
\end{abstract}
\maketitle
\date{}

\section{Introduction \label{sec:intro}}

Since the beginning of the research on the dynamics of complex networks, the deep relationship between topology and dynamics has been thoroughly explored with regard to its effect in the collective state, particularly in
the synchronization between the nodes' dynamics \cite{Pecora1998, Barahona2002, Boccaletti2006, Arenas2008}. A huge effort has been devoted to understand this phenomenon, and the knowledge gathered so far has driven the advances in crucial applications, such as in brain dynamics \cite{Bullmore2009}, power grids \cite{Rohden2012}, and many others where synchronization is {relevant \cite{Pluchino2005,Fujiwara2011}}. In most of them, the focus is placed on a state where all the network units reach the same dynamical state \cite{Arenas2008}. However, there are cases in which the system performs its activity in a partial or weakly { synchronous state  \cite{Rodriguez1999,Lachaux1999,Pecora2014} to preserve} a sort of balance between functional integration and segregation\cite{Tononi1994,Rad2012,Sporns2013}, whereas full synchronization is found to be pathological. As a product of those investigations, it was found that the underlying structure can be inferred from the dynamical correlations among the coupled units in the unsynchronous regime \cite{Arenas2006,Gomez-Gardenes2007,Li2008}. Indeed, the nowadays very active field of functional brain networks relies on the hypothesis that the observed dynamical correlations {are strongly constrained by the anatomical structure \cite{Bullmore2009}, in some cases with a very high correlation between functional and topological networks \cite{Honey2009}.}

% Lachaux, J.-P., Rodriguez, E., Martinerie, J., and Varela, F. J. Measuring phase synchrony in brain signals. Hum. Brain Mapp. 8, 194-208 (1999).
% Honey CJ, Sporns O, Cammoun L, Gigandet X, Thiran JP, Meuli R, Hagmann P (2009) Predicting human resting-state functional connectivity from structural connectivity. Proc Natl Acad Sci U S A 106:2035-2040.
% L. M. Pecora, F. Sorrentino, A. M. Hagerstrom, T. E. Murphy, and R. Roy, Cluster Synchronization and Isolated Desynchronization in Complex Networks with Symmetries. Nat. Commun. 5, 4079 (2014).

It is well known that, in the path to synchrony, the role of the nodes differs as a result of their various topological positions \cite{Gomez-Gardenes2007,Navas2015} as well as of their own intrinsic dynamics \cite{Skardal2014}. Thus, the role of the highly connected nodes (hubs) as coordinators of the dynamics of the whole system has been very often considered \cite{Heuvel2013,Papo2014, Zamora-Lopez2016, Deco2017}. It has been also reported that the hubs are prone to synchronize to each other \cite{Pereira2010} and to the mean field \cite{Zhou2006} in a weakly coupling regime, while the rest of the nodes follow a hierarchical route to synchronization in the process of joining the hubs. 

The fact that synchronization is mediated through the interaction among nodes implies that the single dynamics of each unit is susceptible to change due to the presence of the ensemble. If the connectivity is bidirectional, this perturbation will be stronger the more relevant is the topological position of the node in the network \cite{Zhou2006, Pereira2010}. Therefore, long before the coupling is enough to synchronize the system, each coupled unit is encoding in its own dynamical changes the signature of its role in the structure. In this work, we explore how this relevant feature could be used to extract information  about the network, without having the need to make any reference to pairwise correlations, even in the cases where the structure is unknown. We propose to explore this correlation between the topological rank, measured by the node degree, and the changes in the single node dynamics, measured in terms of its information-based complexity.

\section{Model\label{sec:model}}

 Let us consider a  network of $N$ dynamical units whose $m$ dimensional real state vector $\mathbf{x}_i$ ($i=1,\dots, N$) evolves according to  
\begin{equation}
\dot{\mathbf{x}}_i =  \mathbf{f}(\mathbf{x}_i,\tau_i) - d \sum_j \mathcal{L}_{ij} \mathbf{h}(\mathbf{x}_j),\label{ross}
\end{equation}
where $\mathbf{f}(\mathbf{x}_i,\tau_i)$ is the function governing the node dynamics with $\tau_i$ accounting for some parameter heterogeneity and $d$ is the coupling strength. $\mathcal{L}=\left\lbrace \mathcal{L}_{ij} \right\rbrace$  is the Laplacian matrix describing the coupling structure, with $\mathcal{L}_{ij}= k_i\delta_{ij}-a_{ij}$ where $k_i$ is the node degree, and $\mathbf{A}=\left\lbrace a_{ij} \right\rbrace $ the adjacency matrix, being $a_{ij}=1$ if there is a link between nodes $i,j$ and $a_{ij}=0$ otherwise. The number $N_k$ of nodes having the same degree $k$, is given by the degree probability distribution $P(k)$ as $N_k= NP(k)$.  

In order to address our hypothesis about the relationship between the changes in the { dynamical} properties of each single unit and the number of  neighbors it has, we measure the { Mart\'in-Plastino-Rosso} (MPR) statistical complexity \cite{Lopez-Ruiz1995, Martin2003, Lamberti2004} of ordinal patterns extracted from the signal produced by each dynamical unit, as a function of the node degree $k_i$ and the coupling strength $d$. The methods of analysis of time-series based on {statistical} complexity are gaining relevance in the last years as they provide an easily computable way to quantify the information carried by a signal \cite{Bandt2002, Amigo2018, Politi2017}, and have been applied to a wide variety of systems: from brain data \cite{Amigo2018,Martinez-Huartos2017}, to climate data \cite{Barreiro2011}, or financial analysis \cite{Schnurr2014}. Most of them are based on the {\it permutation entropy}  of the ordinal patterns probability distribution $P_\pi(D)$, where $D$ is the embedding dimension \cite{Bandt2002}. In this study we use the statistical complexity measure defined as $C=H \cdot Q$ \cite{Lopez-Ruiz1995}, where $H = S/S_{max}$ is the normalized permutation entropy, with  $S =-\sum\limits_{\pi} P_{\pi}\log(P_{\pi})$ the Shannon entropy and $S_{max}$ the entropy of the equilibrium probability distribution $P_{e}=1/D!$, and $Q$ is the disequilibrium, measuring { the distance between the two probability distributions} $P_e$ and $P_{\pi}$ by means of the Jensen-Shannon divergence \cite{Zanin2012} (see details in the {\it Appendix}). 

%%%%%%%%%%%%%%% FIGURE 1 %%%%%%%%%%%%%%%%%%%%%%%
\begin{figure}
  \centering
\includegraphics[width=0.8\columnwidth]{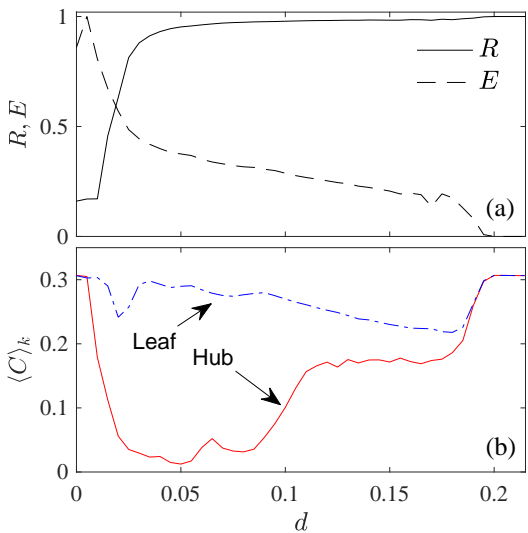}
    \caption{(Color online) (a) Phase order parameter $R$ and synchronization error $E$ vs. $d$ for a star of $N=30$ identical R\"ossler oscillators (see the main text for the parameter values). (b) Dynamical complexity $C$ for the hub and one of the leaves vs. $d$. All the measures are averaged over $10$ initial conditions, storing sequences of maxima of length $10^4$ per node, using $D=4$ as the embedding length for the permutation patterns. \label{fig1}}
\end{figure}
%%%%%%%%%%%%%%%%%%%%%%%%%%%%%%%%%%%%%%%%%%%%%%%%%

In the following sections, we check the generality of our hypotesis in several numerical models (Sec. \ref{sec:NumericalResults}) and provide some theoretical insight (Sec.~\ref{Analytical}) as well as experimental evidence (Sec. \ref{sec:Experimental}). Finally, the results are discussed in Sec. \ref{sec:Conclusions}.

\section{Numerical results \label{sec:NumericalResults}}

\subsection{Chaotic dynamics}

We first check our conjecture by investigating a network of $N$ bidirectionally coupled identical R\"ossler oscillators \cite{Rossler1976} whose time evolution is governed by Eq.~(\ref{ross}), with $\mathbf{x} = (x,y,z)$ as the state vector, ${\mathbf f}(\mathbf{x}) = (-y-z,x+ay,b+z(x-c))$ and ${\mathbf h}(\mathbf{x}) = (0,y,0)$ the vector field and output functions respectively. The coupling strength $d$ is normalized by the maximum node degree of the network $K=\max(k_i)$. The chosen parameters $a=b=0.2$ and $c=9.0$ are such that each R\"ossler unit develops a phase coherent chaotic attractor when isolated. From the time series of the scalar $x_i$ we extract the sequence of $10^4$ maxima. From this last data,  we measure the \textit{amplitude complexity} $C_{i}$ of each node as defined above associated with the probability distribution of all $D!$ permutations $\pi$ of order $D=4$. As we expect that nodes having the same degree $k$ will play equivalent dynamical roles in the ensemble, we compute the evolution of $\langle C\rangle _k$ within a degree class $k$ by averaging over the $N_k$ nodes that have identical degree $k$, i.e., $\langle C \rangle _k= \sum _{[i|k_i=k]}C_i/N_k$.  
  
In addition, we monitor the change in the collective state as the coupling strength $d$ is increased by calculating both the time averaged phase order parameter $R=\frac{1}{N}\langle|\sum_{j=1}^Ne^{{\rm i}\theta_j}|\rangle$, where the phase of the dynamical unit is defined as $\theta_j = \arctan \left(y_j/x_j \right)$, and the time averaged synchronization error $E=\frac{2}{N(N-1)}\langle \sum_{i\ne j} \|{\bf x}_i-{\bf x}_j\|\rangle$, which account for the phase and complete synchronization states, respectively. Throughout the paper, the results are averaged over 10 different networks and initial conditions realizations.

We begin our study with a very simple network configuration, a star of $N=30$ nodes, to grasp the evolution of the dynamical complexity and the role of hubs in heterogeneous networks. In Fig.~\ref{fig1}(a) we report the degree of phase synchronization ($R$, solid line) and the synchronization error ($E$ normalized to its maximum value, dashed line) vs. the coupling strength $d$, observing the two expected transitions that any network of identical phase coherent chaotic oscillators undergo, first a phase synchronization (PS) transition when $R\sim 1$ and later, for larger coupling strength, a complete synchronization (CS) transition with $E=0$. As a star only has two kinds of nodes, $N-1$ leaves and one hub, we plot in Fig.~\ref{fig1}(b) the dynamical complexities $C_i$ of the hub (red solid line) and of one of the leaves (blue dashed-dotted line) as a function of the coupling strength, whose values at $d=0$ coincide as the nodes are identical. For small values of the coupling, when the system is still far from achieving PS, the hub suffers a strong depletion of $C$ which reflects that the leaves are pulling the hub's trajectory out of the original chaotic attractor to a much simpler dynamics, whereas the $C$ value of the leaves remains almost unchanged. As the coupling increases, and the system pass through PS, the $C$ values of leaves and hub get closer until CS into the same original chaotic state is achieved and the initial value of dynamical complexity is recovered. 

%%%%%%%%%%%%%%% FIGURE 2 %%%%%%%%%%%%%%%%%%%%%%%
\begin{figure}
  \centering
 \includegraphics[width=\columnwidth]{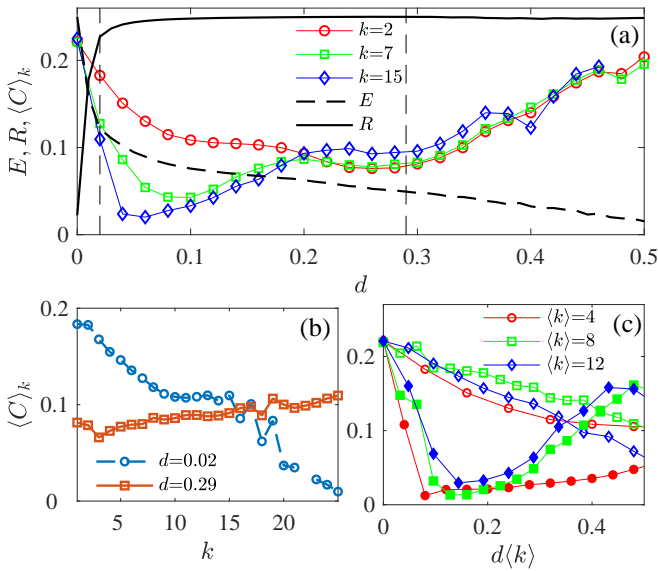}
 \caption{(Color online) Dependence of the dynamical complexity at the node level and its topological role in a SF network of $N=150$ R\"ossler oscillators. (a) Complexity values $C_i$ vs. $d$ for different values of the node degree $k_i$ in a network with $\langle k \rangle=4$. For the sake of comparison, the complete ($E$, black dashed line) and phase ($R$, black continuous line) synchronization curves are shown, rescaled for a better visualization. (b) $\langle C\rangle _k$  vs. $k$ for the two coupling values $d$ marked in (a) with vertical dashed lines, located before ($d=0.02$) and after ($d=0.29$) the phase synchronization transition. (c) $\langle C\rangle _k$ vs. the rescaled coupling $d\langle k\rangle$ for the highest (filled markers) and lowest (void markers) node degree classes for three different mean degrees $\langle k\rangle$ of the networks (see legend). Each point is the average of $10$ network realizations.\label{fig2}}
\end{figure}
%%%%%%%%%%%%%%%%%%%%%%%%%%%%%%%%%%%%%%%%%%%%%%%%

After this preliminary analysis showing a clear dependence of the evolution of the dynamical complexity of each node on its topological role, we check whether this correlation is still observable in more complex topologies. We choose to couple ensembles of $N=150$ R\"ossler oscillators on top of scale-free (SF) networks generated according to \cite{Barabasi2002}, with $\langle k \rangle=4$. In Fig. \ref{fig2}(a) we plot the synchronization measures $E$ (dashed line, values properly rescaled for better comparison)  and $R$ (solid line) along with the $\langle C\rangle _k$ values for several values of $k$. As in the case of the star configuration, there is a clear decrease of $\langle C \rangle _k$ for weak coupling with also a strong hierarchical dependence on $k$ that is lost when the network is clearly phase synchronized. This dependence is much more evident in  Fig.~\ref{fig2}(b) where the $\langle C\rangle _k$ trends for two different coupling regimes are plotted as a function of $k$. At low coupling regime and still far for reaching full PS (vertical dashed line at $d=0.02$ in panel (a)), there is an anti-correlation (blue circles) between $k$ and the dynamical complexity. This behaviour is suggesting an application to structurally rank the nodes in a network according to the complexity of their time series and, therefore, to potentially use this anti-correlation as a proxy for the degree sequence. Note that, at values of the coupling within the full PS regime (vertical dashed line at $d=0.29$ in panel (a)), the dependence is lost, with $\langle C \rangle _k$ almost invariant with $k$.

To further explore the scaling properties of this correlation, we varied the mean degree of the $P(k)$ while preserving the rest of the properties. We found that it scales with $\langle k \rangle$ as shown in Fig.~\ref{fig2}(c) for ensembles of SF networks ($N=150$) with three different mean degrees, where the $\langle C\rangle _k$ is plotted vs. the rescaled coupling $d\langle k\rangle$. It can be seen that the three curves of $\langle C\rangle_k$ for the nodes with the respective highest degree (filled markers) collapse up to exhibiting the same behaviour with $d\langle k\rangle$, as well as those for the nodes with the lowest degree (void markers), whose decreasing trends are much less pronounced. 

\begin{figure}
  \centering
  \includegraphics[width=1\columnwidth,clip]{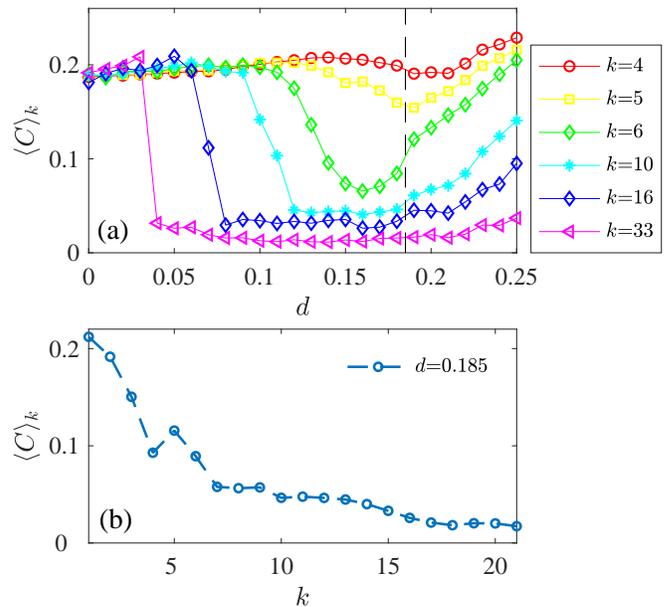}
    \caption{(Color online) Dependence of the dynamical complexity at the node level and its topological role in a SF network of $N=150$  chaotic  Lorenz oscillators and  $\langle k \rangle=4$. (a) Complexity values $\langle C\rangle_k$ vs. $d$ for different values of the node degree $k_i$ (b) $\langle C\rangle _k$  vs. $k$ for the two coupling values $d$=0.17 marked in (a) with a vertical dashed line \label{fig3}}
\end{figure}

In order to test the generality of our results,  we reproduced the study for an ensemble of identical Lorenz oscillators \cite{Lorenz1963} whose chaotic dynamics is far from being phase coherent. In Eqs.~(\ref{ross}), the node dynamics is now replaced by $\mathbf{f(x)} = (10 \cdot (y-x),x\cdot (28-z)-y,xy-(8/3)z)$ and the coupling function is $\mathbf{h}(\mathbf{x}) = (0,y,0)$, with the same network parameters, SF networks of $N=150$ $\langle k\rangle=4$.  Figure~\ref{fig3} shows that the main feature described above is here preserved in this case, with a strong correlation between the dynamical complexity $\langle C\rangle_k$ and the degree, and therefore the possibility to rank the topological relevance of a node only based on individual dynamical measurements.

\begin{figure}
  %\centering
  \includegraphics[width=1.\columnwidth]{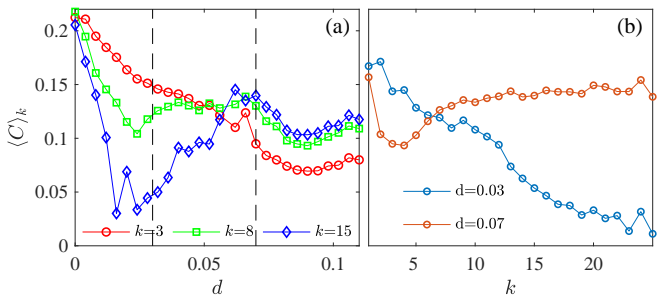}
    \caption{(Color online) Dependence of the dynamical complexity at the node level and its topological role in a SF network of $N=150$ R\"ossler non-identical oscillators with frequency heterogeneity. (a) Complexity values $\langle C\rangle_k$ vs. $d$ for different values of the node degree $k_i$ in a network with $\langle k \rangle=4$.(b) $\langle C\rangle _k$  vs. $k$ for the two coupling values $d$ marked in (a) with vertical dashed lines, located before ($d=0.03$) and after ($d=0.07$) the phase synchronization transition. }\label{fig5}
\end{figure}

The observed negative correlation between $\langle C\rangle_k$ and $k$ featured by networks of chaotic oscillators is not restricted to  ensembles of identical units. In the spirit of evidencing this, the robustness of this relationship is tested by considering an ensemble of slightly different R\"ossler oscillators. In order to do that, we introduce some variability in the R\"ossler natural frequencies considering  ${\mathbf f}(\mathbf{x}) = (-\omega y-z,\omega x+ay,b+z(x-c))$ in Eq.~(\ref{ross}) where the individual node frequencies $\omega$ are set as $\omega_i = 1 \pm \delta \omega_i $ with $\delta \omega_i$ a random value uniformly drawn from the interval $[-0.05, 0.05]$. 
The results are portrayed in Fig.~\ref{fig5}, showing that some level of node heterogeneity does not affect the negative correlation between the dynamical complexity and the node degree. 

\subsection{Stochastic dynamics: The Morris-Lecar neuron}

So far we have considered the node dynamics to be continuous and deterministic, which is a strong limitation in the potential application to real systems with more complicated dynamics and where the presence of intrinsic noise is unavoidable. Therefore, we investigate whether the relationship between structure and dynamics described in previous sections can be extended to stochastic dynamics, in particular to neural dynamics. We implement the bio-inspired Morris-Lecar (ML) model \cite{morris1981} for type II excitatory neurons (with a discontinuous frequency-current response curve), whose equations describing the membrane potential behavior for each unit read \cite{Sancristobal2013, Navas2015}:  

\begin{align}\label{ML}
C \dot{V_i} =  -{}&g_{\rm Ca} M_{\infty}( V_i-V_{\rm Ca} ) - g_{\rm K} W_i( V_i-V_{\rm K} )\nonumber \\ -& g_{\rm l} ( V_i-V_{\rm l}) + q\xi_i + I_i + I^{ext}_i ,  \\
\dot{W_i} = &\phi \,\tau_W  ( W_{\infty}-W_i ) \nonumber
\end{align}

\noindent where $V_i$ and $W_i$ are, respectively, the membrane potential and the fraction of open $\rm K^+$ channels of the $i$th neuron and $M_{\infty}, W_{\infty}$, and $\tau_W$ are hyperbolic functions dependent on $V_i$ and $\phi$ is a reference frequency. The parameters $g_{\rm X}$ and $V_{\rm X}$ account for the electric conductance and equilibrium potentials of the $\rm X=\{K,Ca,\text{leaky}\}$ channels. The external current $I_i^{ext}=50.0$ mA is the same for all the neurons and is chosen such that neurons are sub-threshold to neuronal firing which is induced by the white Gaussian noise $q\xi_i$ of zero mean and intensity $q$. The coupling of the neuron $i$th with the neuron ensemble is described by the injected synaptic current:
\begin{equation}
 I_i= \frac{d}{K} \sum_{j} a_{ij} e^{-2(t-t_j)} (V_0-V_i)
\end{equation}
 given by the superposition of all the post-synaptic potentials emitted by the neighbours of node $i$ in the past, being $t_j$ the time of the last spike of node $j$. The synaptic conductance $d$, normalized by the largest node degree present in the network $K$, plays the role of coupling intensity.

Additionally, the channel voltage-dependent saturation values respond to the dynamics:    
\begin{equation}
M_{\infty}(V_i) = \frac{1}{2}\Bigg[1+\tanh\bigg(\frac{V_i-V_1}{V_2}\bigg)\Bigg], 
\end{equation}

\begin{equation}
W_{\infty}(V_i) = \frac{1}{2}\Bigg[1+\tanh\bigg(\frac{V_i-V_3}{V_4}\bigg)\Bigg],
\end{equation}

\begin{equation}
\tau_W(V_i) = \cosh \bigg( \frac{V_i-V_3}{2V_4} \bigg)
\end{equation}

In Table \ref{tab:param} we detail the values of the parameters used in the simulations, corresponding to type II class excitability for the neuron dynamics which means that a discontinuous transition is found in the dependence of the spiking frequency on the external current.  

\begin{table}
  \caption{Parameters used in the numerical simulations of the Morris-Lecar network in Eqs.~(\ref{ML}).\label{tab:param}}
\begin{ruledtabular}
  \begin{tabular}{ @{\hspace{4em}} lc @{\hspace{4em}}}
 $C$ & $20.0$ $\mu$F/cm$^2$ \\ 
 %\hline
 $g_{\rm Ca}$ & $ 4.0$ $\mu$S/cm$^2$ \\ 
% \hline 
 $g_{\rm K}$ & $ 8.0$ $\mu$S/cm$^2$ \\ 
 %\hline
 $g_{\rm l}$ & $ 2.0$ $\mu$S/cm$^2$ \\   
 %\hline 
 $V_{\rm Ca}$ & $ 120.0 $ mV  \\  
 %\hline
 $V_{\rm K}$ & $-80.0$ mV  \\   
 %\hline
 $V_{\rm l}$ & $ -60.0$ mV  \\   
 %\hline
 $V_1$ & $ -1.2$ mV  \\  
 %\hline 
 $V_2$ & $ 18.0$ mV  \\  
 %\hline 
 $V_3$ & $2.0$ mV\\  
 %\hline 
 $V_4$ & $ 17.4$ mV  \\  
 %\hline 
 $\phi$ & 1/15 \\
 % \hline
\end{tabular}
\end{ruledtabular}
\end{table}

\begin{figure}
  %\centering
 \includegraphics[width=0.4\textwidth]{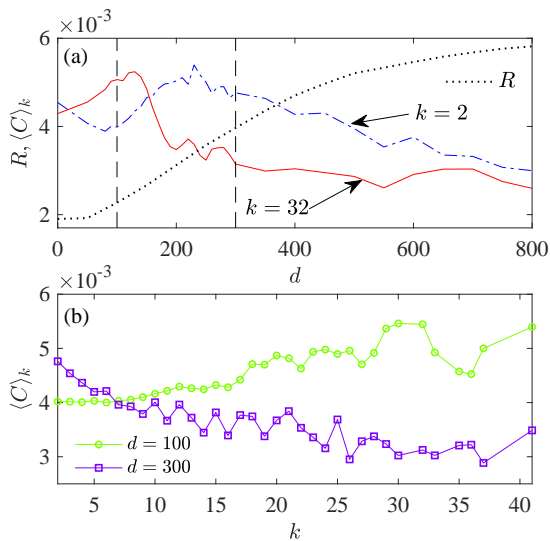}
 \caption{(Color online) Dependence of the dynamical complexity at the node level and its topological role in a SF network of $N=150$ Morris-Lecar neurons. (a) Complexity values $\langle C \rangle_k$ vs. $d$ for low ($k=2$) and high ($k=32$) degree node values in a network with $\langle k \rangle=4$. For the sake of comparison, the phase synchronization curve ($R$, black dotted line) is shown, rescaled for a better visualization. (b) $\langle C \rangle_k$  vs. $k$ for the two coupling values $d$ marked in (a) with vertical dashed lines. Each point is the average of $10$ network realizations.\label{fig4} }
\end{figure}

The typical neuronal dynamics exhibited by Eq.~(\ref{ML}) when $d=0$ consists of a sequence of $L$ spikes produced at random times $t_l$, $l=1,2,\dots,L$, whose amplitude variability is negligible. Therefore, we focused on the complexity $C_i$ of the sequence of inter-spike times $(t_l-t_{l-1})$ patterns of each neuron. Additionally, in order to quantify the level of synchronization, we count how many neurons fire within the same time window \cite{Navas2015}. In order to do this, the total simulation time $T$ is divided in $n=1,\dots,N_b$ bins of a convenient size $\tau$, such that $T=N_b\tau$, and the binary quantity $B_i(n)$ is defined such that $B_i(n)=1$ if the $i$th neuron spiked within the  $n$th interval and $0$ otherwise. The coherence between the spiking sequence of neurons $i$ and $j$ is therefore characterized with the quantity $s_{ij}\in [0,1]$
\begin{equation}
s_{ij}=\frac{\sum_{n=1}^{N_b}B_i(n)B_j(n)}{\sum_{n=1}^{N_b}B_i(n)\sum_{n=1}^{N_b}B_j(n)},
\end{equation}
where the term in the denominator is a normalization factor and $s_{ij}=1$ means full coincidence between the two spiking series. The ensemble average of $s_{ij}$, $S=\langle s_{ij}\rangle =\frac{2}{N(N-1)}\sum_{i,j=1, i\neq j}^N s_{ij}$ is conveniently rescaled and reported in Fig.~\ref{fig4} as a dotted line indicating a transition from an asynchronous to an almost synchronous firing as the synaptic conductance $d$ is increased. Superimposed to this curve are the complexities $C_i$ of nodes with low ($k=2$, blue dash-dotted line) and high ($k=32$, red solid line) degrees for $10$ realizations of a SF network of $N=150$ ML neurons. % with $\langle k\rangle =4$ neighbors in average. 

%%%%%%%%%%%%%%% FIGURE 4 %%%%%%%%%%%%%%%%%%%%%%%

We observe that, as the coupling increases, the complexity of the highly connected nodes peaks at incipient levels of synchronization, as well as for the low degree nodes - which occurs later. This is due to the fact, for small coupling values, the hubs are cross-talking with many nodes receiving incoherent, noise-induced signals contributing to increase its own complexity. For larger values of the coupling strength ($d=300$), still far from PS, the hubs complexity decreases as the increase of inputs pushes the neuron towards  the periodic transition. In the bottom panel of Fig.~\ref{fig4}, we show the correlation between the complexity $\langle C \rangle_k$ and the node degree at the two coupling strengths marked with dashed lines in the upper plot. Again, as in the case of deterministic dynamics, a negative correlation of the complexity values with the number of synapses appears for intermediate values of the synchronization level. This suggests that, indeed, there is a region close to full synchronization where the complexity of a node can tell us about its degree.

\section{An analytical insight \label{Analytical}}
 
The behavior can be understood analytically by performing a mean field approximation and a linear stability analysis of the actual state of each oscillator in the weakly coupling regime where the system is still far from reaching the same collective state \cite{Zhou2006,Pereira2010}. The local mean field that oscillator $i$ is receiving is $\bar{\mathbf x}_i=k_i^{-1}\sum_{j=1}^{N} a_{ij} {\mathbf h}({\mathbf x}_j)$. In the case of a highly connected node ($k_i\gg 1$), its mean field can be well approximated by the global mean field ${\mathbf X}=N^{-1}\sum_{j=1}^N {\mathbf h}({\mathbf x}_j)$, that is $\bar{\mathbf{x}}_i\sim {\mathbf X}$, whose variance is, below the onset of synchronization, very small \cite{Rosenblum2004,Pereira2010}. Under this assumption, the contribution from the coupling term to the time evolution of the hubs in Eq.~(\ref{ross}) is simply $N {\mathbf X}$ that can be neglected since it is either zero or a constant depending whether the attractor has a symmetry with respect to the origin.  Therefore, the governing equations for the hubs \cite{Pereira2010} are given by
\begin{equation}
\dot{\mathbf{x}_i} \simeq	{\mathbf f}(\mathbf{x}_i) + dN{\mathbf X} - d k_i {\mathbf h}(\mathbf{x}_i) \label{weakly}
\end{equation}
 
\noindent that is, the hub's dynamics is being modulated by a strong negative self-feedback term ($-dk_i{\mathbf h}({\mathbf x}_i))$ that stabilizes the unstable periodic orbits resulting in a more stable trajectory than the original uncoupled one \cite{Boccaletti2000}. To prove this, let us consider all the infinitesimal displacements $\delta{\mathbf x}$ from a given trajectory ${\mathbf x}_i$ of a hub. The time evolution of the tangent vector $\delta{\mathbf x}_i$ is given by the linearization of the Eq.~(\ref{weakly}): 
\begin{equation}
\delta{\dot{\mathbf x}_i}=\left[{\mathbf J}{\mathbf f}({\mathbf x}_i)-d k_i {\mathbf J}{\mathbf h}({\mathbf x}_i)\right] \delta {\mathbf x}_i\label{linear}
\end{equation} 
where ${\mathbf J}$ stands for the Jacobian matrix. %If the position of the attractor is independent of the trajectory, then 
Without loss of generality, assuming that the coupling function ${\bf h}$ is linear ,the solution to the variational equations of the perturbations results in an exponentially growth at a rate given by the Lyapunov exponents, whose maximum is given by $\Lambda(k)=\Lambda_0 - dk_i $ where $\Lambda_0$ is the maximum positive Lyapunov exponent corresponding to a chaotic uncoupled oscillator. %On the other hand, the rate at which the perturbation dumps out is related to the Kolmogorov-Sinai entropy \cite{Pesin1976,Bandt2002} in the sense that the entropy accounts for the total expansion of the system. Therefore, the entropy $H_K(k)$ of the node with degree $k$ can be estimated as $H_K(k)=H_K(0)-d k$, being $H_K(0)$ the value for the uncoupled system. %
As a consequence, the trajectory will become dynamically less complex as a linear function of $k$, as observed in Fig. \ref{fig2}(b). Eventually, if the original node is chaotic and highly connected, it can become periodic with the consequent loss of statistical complexity. On the contrary, for the less connected nodes $k_i\sim 1$, in the weakly coupling regime, the diffusive term is too small as to modify the trajectory, and the node dynamics retains most of its original complexity.

\section{ Experimental implementation \label{sec:Experimental}}

\begin{figure}
	\centering
\includegraphics[width=0.5\textwidth]{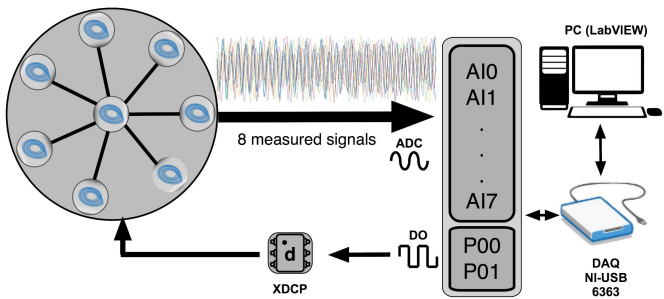}
		\caption{Schematic representation of the experimental setup of a star network with eight R\"{o}ssler-like oscillators. Using ADCs, signals are measured and stored by a DAQ card and a PC, while DO ports change the value of the coupling strength $d$ through a digital potentiometer (XDCP).}
	\label{fig6}
\end{figure}
In order to provide some experimental evidence, we designed a star network with eight bidirectionally coupled R\"ossler-like chaotic electronic circuits. We implement a setup consisting on an electronic version of the R\"{o}ssler-like system \cite{Carroll1995} described by the following equations:

\begin{align}
\label{equ:LtR}
\dot{v}_{1i}&=-\frac{1}{R_{1}C_{1}}\left(v_{1i} + \frac{R_{1}}{R_{2}}v_{2i}+\frac{R_{1}}{R_{4}}v_{3i}\right)\\\nonumber
\dot{v}_{2i}&=-\frac{1}{R_{6}C_{2}}\left[-\frac{R_{6}R_{8}}{R_{9}R_{7}}v_{1i}+ \left(1- \frac{R_{6}R_{8}}{R_{c}R_{7}}\right)v_{2i}\right]+\\
&+\frac{d}{R_{15}C_2} \sum_{j=1}^{N}a_{ij}\left[{v_{2j}-v_{2i}}\right]   \\
\dot{v}_{3i}&=-\frac{1}{R_{10}C_{3}}\left(-\frac{R_{10}}{R_{11}}G_{ v_{1i}}+v_{3i} \right),
\end{align}
and the piecewise function $G_{v_{1i}}$ as
\begin{align}\label{LtR2}
G_{v_{1i}}= \left\{ \begin{array}{lcc}
0 	&   \text{if}  & v_{1i} \le v_{\rm ref} \\
\\ G_0  &   \text{if}  & v_{1i}> v_{\rm ref} \\
\end{array}\right.
\end{align}

\noindent where $v_{\rm ref} = {V_d}\Big( 1+ \frac{R_{14}}{R_{13}}\Big)+V_{ee}\frac{R_{14}}{R_{13}}$ and $ G_0 = \frac{R_{12}}{R_{14}}v_{1i}-V_{ee}\frac{R_{12}}{R_{13}}-{V_d}\left(\frac{R_{12}}{R_{13}}+\frac{R_{12}}{R_{14}} \right)$. All parameter values are listed in Table~\ref{table:components} and a schematic representation of the experimental setup is shown in Fig.~\ref{fig6}.
%The details of the circuit are shown in %Table~\ref{}, where a full description of every component’s parameters is given.% (also, $V_{ee}=15$).
We refer the interested reader to Ref.~\cite{Datainbrief} for the visualization of the electronic and coupling circuits and to Refs.~\cite{Sevilla2015,Sevilla2016,Leyva2017,Leyva2018} for a detailed description of the experimental implementation of the circuits and previous realizations in different network configurations. The Analog-to-Digital Cards (ADCs) (AI0...AI7) ports from the Data Acquisition (DAQ) card are used for sampling the variable $v_{2}$ of each circuit. A coupler is introduced between the circuits. The coupling circuit is based on a differential operational amplifier (Op-Amp) where the $v_{2j}$ and $v_{2i}$ signals are introduced. A digital potentiometer (XDCP) is used to vary the gain of the amplifier, which is adjusted by digital pulses from digital ports (DO).  Here P00 is used to increase or decrease the resistance of the voltage divisor ($d$), while P01 sets the value of the resistance (100 discretized steps, 1 step = 100$\Omega$). The entire experimental process is controlled by a virtual interface in LabVIEW 2016 (PC).

\begin{table}
  	\caption{Values of the electronic components and constants used for the construction of the R\"{o}ssler-like oscillator.
	\label{table:components}}
  \begin{ruledtabular}
    % \begin{tabular}{ @{\hspace{3em}} l l @{\hspace{3em}} l @{\hspace{3em}} l }
	\begin{tabular}{l l l l }
%	    \hline
		$C_{1}=1$nF				& $C_{2}=1$nF				& $C_{3}=1$nF				& {$V_{ee}=15$ V} \\
%		\hline
		$R_{1}=2$\,M$\Omega$ 	& $R_{2}=200$\,k$\Omega$& $R_{3}=10$\,k$\Omega$ & $R_{4}=100\,k\Omega$ \\
%		\hline
		$R_{5}=50$\,k$\Omega$ & $R_{6}=\,5$M$\Omega$ 	& $R_{7}=100$\,k$\Omega$& $R_{8}=10$\,k$\Omega$ \\
%		\hline
		$R_{9}=10$\,k$\Omega$ 	& $R_{10}=100$\,k$\Omega$ & $R_{11}=100$\,k$\Omega$& $R_{12}=150$\,k$\Omega$ \\
%		\hline
		$R_{13}=68$\,k$\Omega$ & $R_{14}=10$\,k$\Omega$ & $R_{15}=100$\,k$\Omega$& {$V_{d}=0.7$ V}   \\
%		\hline
	\end{tabular}
    \end{ruledtabular}
\end{table}

%%%%%%%%%%%%%%% FIGURE 4 %%%%%%%%%%%%%%%%%%%%%%%
\begin{figure}[b]
  \centering
  \includegraphics[width=0.35\textwidth]{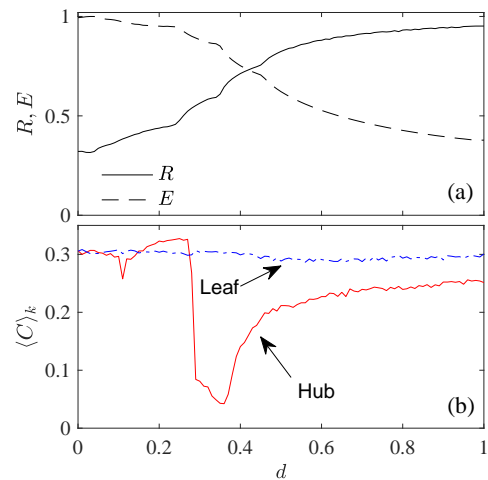}
 \caption{(a) Phase order parameter $R$ and synchronization error $E$ vs. $d$ for a star of $N=8$ almost identical (within the $5\%$ experimental tolerance) R\"ossler-like electronic circuits for the system experimental description. (b) Dynamical complexity of the hub and of one of the leaves vs. $d$. Complexity measures are averaged over 30 different initial conditions and calculated over sequences of 5000 maxima with embedding length $D=3$.\label{fig7}}
\end{figure}
%%%%%%%%%%%%%%%%%%%%%%%%%%%%%%%%%%%%%%%%%%%%%%%%%
The experiment works in the following way: first, $d$ is set to zero and digital pulses (P00 and P01) are sent to the digital potentiometer (X9C103) until the value of maximum resistance is reached. After waiting $500$ ms, $d$ is varied from 0 to 10 k$\Omega$ in 100 steps and at each step, the measure is repeated for thirty different initial conditions. For each coupling value, the variables $v_{2i}$ of the circuits are acquired by the analog ports (AI0...AI7), and the synchronization error is calculated and stored in the PC. The local maxima of each oscillator (5000 maxima) are located and stored to perform the corresponding complexity measures, as explained in Sec. \ref{sec:model}.

The results are presented in Fig.~\ref{fig7}, where the synchronization state (Fig. \ref{fig7}(a)) and dynamical complexity of the hub and of one of the leaves (Fig.~\ref{fig7}(b)) are to be compared with their numerical counterparts in Figs.~\ref{fig1}(a)-(b). Despite the natural parameter mismatch and environmental noise affecting our experimental setup, the two markedly different paths of the dynamical complexities of both the hub and the leaves, a large loss in the hub and an almost constant level of complexity in the leaves, largely agree with those obtained in the numerical simulation, confirming the generality of the observation.

\section{Conclusions \label{sec:Conclusions}}

In this work, we have inspected the relationship between the topological role of a node in a complex network and its dynamical behavior, represented by its  complexity. We show, both numerically and experimentally, that in a simple star of identical chaotic oscillators, the hub exhibits a minimum of complexity in the route to synchronization while the leaves almost keep unperturbed their initial complex behavior. When considering more heterogeneous degree distributions, the same behavior is observed in the route to synchronization, with higher degree nodes exhibiting lower values of complexity. Importantly, when comparing the  complexity of each node and its degree, we found a distinctive linear correlation with higher degree nodes exhibiting less  complexity and that is generally observed in networks of other types of chaotic oscillators or pulse-coupled neurons. The reported results  could explain recent observations about the low complexity of the hubs in functional brain networks \cite{Martinez-Huartos2017} but, beyond than that, they suggest that the role played by the topology of a network could be unveiled by just computing the dynamical complexity associated with the time series sampled at each node. The fact that structural information of a network can be inferred without computing pairwise correlations like those commonly performed in functional networks could be exploited in diverse fields as neuroscience, econophysics or power grids. 

%For various types of nets, we have given a detailed description of why this interplay exists and what it intuitively means.

\begin{acknowledgments}
Financial support  from the Ministerio de Econom\'ia y Competitividad
of Spain (projects FIS2013-41057-P and FIS2017-84151-P) and from the Group of Research 
Excelence URJC-Banco de Santander is acknowledged. We thank J.M. Buld\'u for fruitful discussions. R.S.E. acknowledges support from Consejo Nacional de Ciencia y Tecnolog\'ia call SEP-CONACYT/CB-2016-01, grant number 285909.
\end{acknowledgments}

\appendix*
\section{Ordinal patterns and complexity measure \label{appendix}}

The ordinal patterns formalism  \cite{Bandt2002} associates a symbolic sequence to a time series, transforming the actual values of the measure into a set of natural numbers. For doing that, the time series is divided in bins of  size $D$. In each bin, the $D$ data values are ordered in terms of its relative amplitudes   \cite{Rad2012}, which provides the correspondent symbolic sequence. The information content of these sequences is then evaluated as a function of the complexity measure. This is a broad-field, well-established and known method, statistically reliable and robust to noise, extremely fast in computation and with a clear definition and interpretation in physical terms. It is derived from two also well-established measures (divergence and entropy), also easily interpretable when analyzing non linear dynamical systems. In addition, it only requires soft criteria, namely that the time series must be pseudo-stationary and that $M>>D!$ (where $M$ is the number of points of the entire time series), which are easily checkable. We proceed in the following way:

\begin{enumerate}
\item We count how many times a certain symbolic order sequence (or {\it pattern}) of size $D$ appears ($N_\pi$).
\item We then define a probability of occurrence for each pattern: $P_\pi = \frac{N_\pi}{N_T}$, where $N_T$ is the total number of patterns in which we divide the time series, i.e. $N_T = N/D$.
\item We construct an \textit{empirical probability distribution}, which we call $P$ from now on, from the pool of $P_\pi$.
\end{enumerate}

Once it is obtained the probability distribution $P$, we can now define the dynamical complexity, a measure that should be minimal both for pure noise and absolute regularity, and provide a bounded value for other regimes. Being this so, we need to characterize the disorder and a correcting term (i.e., a way of comparing known probability distributions with the actual one). In the main text, we define the dynamical complexity ($C=H Q$) as the product of the Permutation Entropy ($H$) and the Disequilibrium ($Q$).

To define the permutation entropy $H$, the first step is the evaluation of the Shannon entropy, that gives an idea of the \textit{disorder} of the series:
\begin{equation}
S[P] = - \sum_{j=1}^{D!} p_j \cdot \log(p_j)
\end{equation}
The permutation entropy corresponds to the normalization of $S$ respect to the entropy of the uniform probability distribution, $S_{max}$:
\begin{align}
&H = \frac{S}{S_{max}}, \quad S_{max} = S[P_e], \\
& P_e \equiv \lbrace 1/D! \rbrace_{1,...,D!}  \implies 0 \leq H \leq 1 \nonumber
\end{align}

Regarding the disequilibrium $Q$, it is a way of measuring the distance of the actual probability distribution $P$ with the equilibrium probability distribution $P_e$. This notion of {\it distance} can be acquired by several means; in this text, we adopt the statistical distance given by the Kullback-Leibler \cite{Kullback1951} relative entropy ($K$):
\begin{align}\label{eq:KL} \nonumber
K[P|P_e] &= - \sum_{j=1}^{D!} p_j \cdot \log(p_e) + \sum_{j=1}^{D!} p_j \cdot \log(p_j) = \\
&= S[P|P_e] - S[P] 
\end{align}

\noindent where $S[P|P_e]$ is the Shannon cross entropy. If we now make symmetric Eq.~ (\ref{eq:KL}), we get the Jensen-Shannon divergence ($J$):

\begin{equation}\label{eq:J}
J[P|P_e] = (K[P|P_e]+K[P_e|P])/2
\end{equation}

For our purposes, it is highly convenient to write (\ref{eq:J}) in terms of $S$ solely:

\begin{equation}
J[P|P_e] = S[(P+P_e)/2] - S[P]/2 - S[P_e]/2
\end{equation}
Finally, we can write the disequilibrium $Q$ as the normalized version of $J$ as:
\begin{equation}
Q = Q_0 J[P|P_e]
\end{equation}
with $Q_0 = {\frac{N+1}{N}\log(N+1)-2\log(2N)+\log(N)}^{-1}$, implying again $0 \leq Q \leq 1$.

\bibliography{references}

\end{document}